\documentclass[prd,nofootinbib, amssymb,twocolumn,preprintnumbers]{revtex4-1}
\usepackage{graphicx, epsfig, bm, amsmath}
\usepackage{amsmath}
\usepackage{amssymb}
\usepackage{url}
\usepackage{subfigure}
\usepackage{textcomp}
\usepackage{graphicx}
\usepackage{bm}
\usepackage{dcolumn}
\usepackage[usenames]{color}

\newcommand{\be}{\begin{equation}}
\newcommand{\ee}{\end{equation}}
\newcommand{\bea}{\begin{eqnarray}}
\newcommand{\eea}{\end{eqnarray}}

\newcommand{\bse}{\begin{subequations}}
\newcommand{\ese}{\end{subequations}}
\newcommand{\bi}{\begin{itemize}}
\newcommand{\ei}{\end{itemize}}

\def\beq{\begin{equation}}
\def\eeq{\end{equation}}
\def\bea{\begin{eqnarray}}
\def\eea{\end{eqnarray}}

\begin{document}

\preprint{ IPM/P-2012/052}  

\vskip1cm

\title{Slow-roll trajectories in Chromo-Natural and Gauge-flation Models,\\ an exhaustive analysis}

\author{Azadeh Maleknejad $^{1}$}
\email[]{azade[AT]ipm.ir}

\author{Moslem Zarei$^{1,2}$}

\email[]{m.zarei[AT]cc.iut.ac.ir}

\affiliation{$^{1}$ School of Physics, Institute for Research in Fundamental Sciences (IPM), P. O. Box 19395-5531, Tehran, Iran}

\affiliation{$^{2}$Department of Physics, Isfahan University of Technology, Isfahan 84156-83111, Iran}


\begin{abstract}

We present an exhaustive analysis on the background inflationary solutions of the chromo-natural model. We show that starting from an arbitrary axion field value $\chi_0\in(0,f\pi)$, it is possible to have slow-roll inflation with enough number of e-folds and determine the allowed region of the parameters corresponding to each $\chi_0$ value. Having the available parameter space, we then study the behavior of the solutions with respect to the initial value of the axion field.

\end{abstract}


\maketitle

\section{ Introduction}

Inflation paradigm is the main framework for the early universe cosmology today which solves the old cosmological puzzles such as horizon and flatness problems as well as providing a natural mechanism for generating cosmic perturbations. Despite of its many phenomenological successes, however, inflation faces some theoretical challenges. Most of the successful inflationary models in the literature are based on using one or more scalar fields with a sufficiently flat potential to guarantee the slow-roll condition and leads to enough number of e-foldings. On the other hand, without providing a consistent high energy theory setting, such extremely flat potentials generally is not protected against the quantum corrections unless we highly fine-tuned the system. This theoretical concern, the so-called $\eta$-problem, has driven many efforts to avoid fine-tuning in the potential. First used in Natural Inflation \cite{Freese:1990rb,Adams:1992bn}, one possible solution is considering axion fields as inflaton. In fact, since the axion field has a perturbatively exact cosine-type potential, it would not received any perturbative quantum corrections. However, in order to have a successful slow-roll inflation, the axion decay constant f should be very large in this model ($f\sim M_{pl}$), which is not a natural scale in beyond standard model and string theory
motivated particle physics models \cite{Freese:2004un}. The interesting idea of using gauge fields to fix this naturalness problem, first proposed in \cite{Anber:2009ua}, where considering the quantum back reaction of a U(1) gauge field on the axion condensate, system leads to a phase of slow-roll inflation even with a steep axion potential.
Recently, a novel inflationary model, chromo-natural, has been proposed in which considering non-Abelian gauge fields and their interaction with the axion at the background level, solves the above problem \cite{Adshead:2012kp,Adshead:2012qe}. For an extensive review on this topic see \cite{Maleknejad:2012fw}.

In the chromo-natural model, the Chern-Simons interaction term between the axion and the non-Abelian gauge field plays an important role. By converting most of the potential energy of the axion field into gauge field energy, this term slows down the evolution of the axion field. This fact then makes it possible to have successful slow-roll inflation even with the natural axion decay constants $(f\ll M_{pl})$.
The potential of the axion field $V=\mu^4(1+\cos(\chi/f))$, has a maximum at $\chi/f=0$ and a minimum at $\chi/f=\pi$. Starting the slow-roll inflation from a small axion field initial value ($\chi_0/f\ll 1$), we have a hilltop-type model corresponds to the maximum of axion potential. On the other hand, if one considers $\chi_0/f$ values close to $\pi$, this leads to a slow-roll inflationary model corresponds to the minimum of the axion potential. The former is the model discussed in \cite{Adshead:2012kp,Adshead:2012qe}, while the later is effectively the gauge-flation model \cite{Maleknejad:2011jw,Maleknejad:2011sq,SheikhJabbari:2012qf}.

However, the chromo-natural model has a much richer parameter space and possible solutions.
In fact, starting from any arbitrary initial axion field value $\chi_0\in(0,f\pi)$, there is a slow-roll trajectory in this model. An exhaustive analysis on the classical solutions of the chromo-natural model is the main focus of this paper. Here, working out the analytical solution of number of e-folding in terms of the parameters of the system and $\chi_0$, we determine the space of parameters which leads to a well controlled slow-roll inflation corresponding to the initial value of the axion field $\chi_0$. After that, we study the system numerically in different regions of the parameter space, and illustrate its behavior with respect to the initial value of the axion field.

This paper is organized as follows. In section \ref{Setup}, we review the chromo-natural model and its slow-roll inflationary trajectories. In section \ref{parameter space}, first we work out the exact analytical from of number of e-folding in terms of the other parameters of the system. Then we present the exhaustive parameter space analysis of the model which leads to successful inflationary solutions. In section \ref{numerical}, we study the system numerically and illustrate the generic behavior of the system in different initial axion field values. Finally, we conclude in \ref{conclusion}.

Throughout this work, we use natural units where the reduced Planck mass $M_{pl}=(8\pi G)^{-1/2}=1$.

\section{Setup}\label{Set up}
In \cite{Adshead:2012kp}, an inflationary model consisting an axion field $\chi$ and a non-Abelian $SU(2)$ gauge field $A^a_{~\mu}$ has been proposed
\bea\label{model}
S&=&\int d^4x\sqrt{-\verb"g"}(-\frac{R}{2}-\frac14F^a_{~\mu\nu}F_a^{~\mu\nu}-\frac12(\partial_\mu\chi)^2\nonumber\\
&~~~&-\mu^4(1+\cos(\frac{\chi}{f}))+\frac{\lambda}{8f}\chi(F\wedge F)),
\eea
where $F\wedge F=\epsilon^{\mu\nu\lambda\sigma}F^a_{~\mu\nu}F^a_{~\lambda\sigma}
$, $e^{0123}\equiv+1$ and $F^a_{~\mu\nu}$ is the gauge field strength tensor, given as
\be
F^a_{~\mu\nu}=\partial_\mu A^a_{~\nu}-\partial_\nu A^a_{~\mu}-g\epsilon^a_{~bc}A^a_{~\mu}A^b_{~\nu}.
\ee
Here $a,b,c=1,2,3$ represent indices of algebra while $\mu,\nu=0,1,2,3$ are used for the space-time indices.
The action \eqref{model} describes a four parameter model, dimensionless parameters $g,\lambda$ and dimensionful $\mu$ and $f$.

Respecting the strong energy condition, the energy-momentum tensor  of the Yang-Mills term is traceless, while the $F\wedge F$ is a topological term which does not contribute to the energy-momentum tensor. As a result, here the gauge field alone does not lead to an inflationary expansion and system inflates only in case that the axion potential is the dominate part of the energy density. On the other hand, in the absence of the gauge field-axion interaction, successful slow-roll inflation requires a super-Planckian axion decay constant $f\sim M_{pl}$ \cite{Freese:2004un}, which is not a natural scale to realize in string theory \cite{Banks:2003sx}. In fact, turning on the non-Abelian gauge field, $A^a_{~\mu}$ in the background, the gauge field-axion interaction term slows down the axion's evolution and makes it possible to have successful slow-roll even with sub-Planckian natural $f$ values \cite{Adshead:2012qe}.

Since we are interested in the isotropic and homogeneous FRW cosmology
\be\label{FRW}
ds^2=-dt^2+a^2(t)\delta_{ij}dx^idx^j,
\ee
the temporal gauge $A^a_{~0}=0$ appears to be a suitable gauge fixing. As discussed in details in \cite{Maleknejad:2011jw,Maleknejad:2011sq}, this fixes the gauge symmetry up to
time independent SU(2) gauge transformations which is the key to restoring
the rotation symmetry in spite of the background non-Abelain gauge fields. Thus, we obtain the following homogeneous isotropic ansatz \cite{Maleknejad:2011jw,Maleknejad:2011sq,homo-iso-gauge-config-2}
\be\label{A-ansatz-background}
A^a_{~\mu}=\left\{
\begin{array}{ll} \phi(t)\delta^a_i\, ,\qquad  &\mu=i
\\   0\,, \qquad &\mu=0\,.
\end{array}\right.
\ee%
where $\phi(t)$ is not a scalar under rotation transformations, while%
\be\label{psi-def}%
\psi(t)=\frac{\phi(t)}{a(t)},%
\ee%
is indeed a scalar. (Note that the spatial triads of the metric \eqref{FRW} can be chosen as
$e^a_{~i}=a(t)\delta^a_i$.)
In fact, due to the above property, all non-Abelian gauge field theories can provide the
setting for constructing an isotropic and homogeneous inflationary background \cite{Maleknejad:2011jw,Maleknejad:2011sq}.

After plugging \eqref{FRW} and \eqref{A-ansatz-background} into \eqref{model}, we obtain the the following reduced Lagrangian for the matter sector
\bea
\label{redL}
\mathcal{L}_{\textit{red.}}&=&\frac32(\frac{\dot\phi^2}{a^2}-\frac{g^2\phi^4}{a^4})+\frac12\dot\chi^2
-\mu^4(1+\cos(\frac{\chi}{f}))\nonumber\\
&-&3\frac{g\lambda\chi}{f}\frac{\dot\phi\phi^2}{a^3}.
\eea
One can show \cite{Maleknejad:2011jw,Maleknejad:2011sq} that the reduction to the ``isotropic sector'' is consistent. 
Thus the Friedman equations governing the evolution of system are
\bea\label{H2}
H^2&=&\frac12\left(\frac{\dot\phi^2}{a^2}+\frac{g^2\phi^4}{a^4}\right)+\frac{\dot\chi^2}{6}+\frac{\mu^4}{3}(1+\cos(\frac{\chi}{f})),~~~\\
\label{dH}
\dot{H}&=&-\left(\frac{\dot\phi^2}{a^2}+\frac{g^2\phi^4}{a^4}+\frac12\dot\chi^2\right).
\eea

We consider slow-roll inflation, during which the slow-roll parameters defined below are very small
\be\label{slow-roll-par}
\epsilon\equiv-\frac{\dot H}{H^2},\quad \eta\equiv-\frac{\ddot{H}}{2\dot{H}H}.
\ee
Then, combining the slow-roll conditions with \eqref{H2} and \eqref{dH} gives the following approximation for $H^2$ during inflation
\be\label{H2-sl}
H^2\simeq\frac13\mu^4(1+\cos(\frac{\chi}{f})),
\ee
where $\simeq$ means up to the first order in $\epsilon$.
One can determine $\epsilon$
\be\label{epsilon-dec}
\epsilon=\epsilon_\psi+\epsilon_\chi,
\ee
here $\epsilon_\psi$ denotes the gauge-field contribution and $\epsilon_\chi$ represents the axion contribution to $\epsilon$ and given as
\be
\epsilon_\psi=\left((1-\delta)^2+\gamma\right)\psi^2,\quad \epsilon_\chi=\frac12\frac{\dot\chi^2}{H^2},
\ee
where 
\be
\delta\equiv-\frac{\dot\psi}{H\psi},\quad \gamma\equiv\frac{g^2\psi^2}{H^2}.
\ee

Since both of them are positive, slow-roll condition demands that each of $\epsilon_\psi$ and $\epsilon_\chi$ be very small during inflation.
Also combing \eqref{slow-roll-par} and \eqref{epsilon-dec}, we can determine $\eta$
\be
\eta=\eta_\chi+\eta_\psi,
\ee
where again $\eta_\chi$, $\eta_\psi$ are $\chi$ and $\psi$ contributions to $\eta$ respectively
\bea
\eta_\chi&=&-\frac{\ddot\chi}{H\dot\chi}\frac{\epsilon_\chi}{\epsilon },\\
\label{eta-psi}
\eta_\psi&=&\left(\frac{2\delta}{\epsilon}+\frac{(1-\delta)^2}{((1-\delta)^2+\gamma)}(\frac{\frac{\dot\delta}{\epsilon H}}{1-\delta}+1-\frac{\delta}{\epsilon})\right)\epsilon_\psi.~~~
\eea
Consequently, a successful slow-roll inflation requires that $\epsilon_\psi$, $\epsilon_\chi$ and $\eta_\chi$ be very small, we should also demand $\delta\lesssim\epsilon$. As we will see shortly, dependending on the value of $\frac{\chi_0}{f}$, $\delta$ should be a quantity of the order $\epsilon^2$ to $\epsilon$.

Up to this point, using Friedman equations, we obtained the slow-roll parameters. Next, we study the background trajectories of $\psi$ and $\chi$ fields.
From the action \eqref{model}, we have the field equations of $\psi$ and $\chi$ respectively as below
 \bea
 \label{chieq}
 &~&\ddot{\chi}+3H\dot{\chi}-\frac{\mu^4}{f}\sin(\frac{\chi}{f})=-3g\frac{\lambda}{f}\psi^2(H\psi+\dot\psi),\\
 \label{psieq}
 &~&\ddot{\psi}+3H\dot\psi+(\dot H+2H^2)\psi+2g^2\psi^3=g\frac{\lambda}{f}\psi^2\dot\chi.
 \eea
Then, imposing the slow-roll conditions in \eqref{chieq}, we obtain the attractor trajectory of $\psi$
\be\label{psisl}
\psi_{min}\simeq(\frac{\mu^4\sin(\frac{\chi}{f})}{3g\lambda H})^{1/3}.
\ee
Note that, one can also derive the above result by minimizing the effective potential of the axion in \eqref{redL}
\be\label{U-eff}
U_{eff}(\chi)=\mu^4(1+\cos(\frac{\chi}{f}))+\frac{3g\lambda\chi}{f}\frac{\dot\phi\phi^2}{a^3},
\ee
which up to the dominant order in slow-roll leads to \eqref{psisl} for $\psi$ \cite{SheikhJabbari:2012qf, Adshead:2012kp}.
Plugging \eqref{H2-sl} into \eqref{psisl}, we then obtain the following simple form for $\psi$ during the slow-roll
\be\label{psi-sl}
\psi_{min}\simeq(\frac{\Upsilon}{\lambda})^{1/2}(\sin(\frac{\chi}{2f}))^{1/3},
\ee
where 
\be
\Upsilon\equiv(\frac{2\lambda\mu^4}{3g^2})^\frac{1}{3},
 \ee
 and it leads to $\delta\propto\cos(\frac{\chi}{2f})$. On the other hand, \eqref{eta-psi} indicates that for having slow-roll inflation we need $\delta\lesssim\epsilon$. Thus, in cases with small $\frac{\chi}{f}$ values, $\delta$ should be of the order $\epsilon$, while for cases with $\frac{\chi}{f}=\pi+\mathcal{O}(\epsilon)$, we have $\delta\sim\epsilon^2$ and for the generic values of $\chi_0$, $\delta$ should be somewhere between $\epsilon^2$ and $\epsilon$. The former hilltop-type model is discussed in \cite{Adshead:2012kp}, while as shown in \cite{SheikhJabbari:2012qf}, the latter limit corresponds to gauge-flation \cite{Maleknejad:2012fw,Maleknejad:2011jw}. In Fig. \ref{delta-chi} we see that solutions with larger initial axion field values experience smaller $\psi$ field evolutions during the slow-roll inflation, and hence smaller $\delta$ values, which confirms our analysis.

Writing \eqref{psieq} up to the leading terms in slow-roll and after using  \eqref{psisl}, we obtain
\be
\epsilon_\psi\simeq\frac12\frac{g\lambda}{fH}\psi^3\frac{\dot\chi}{H}.
\ee
On the other hand, during the slow-roll, \eqref{chieq} implies that $\frac{\dot\chi}{H}\ll\frac{g\lambda}{fH}\psi^3$ which leads to
\be
\epsilon_\chi\ll\epsilon_\psi,
\ee
and as a result, we have
\be\label{eps}
\epsilon\simeq\epsilon_\psi.
\ee
Thus, during the slow-roll inflation, the potential energy of the axion field is mostly converted into the gauge field energy rather than the kinetic energy of axion itself.

Finally, using \eqref{psisl} and \eqref{eps}, one can compute the number of e-foldings, given by the following integral
\be\label{N}
N\simeq\frac{\lambda\Upsilon}{\sqrt[3]{4}}\int^\pi _{\frac{\chi_0}{f}}\frac{(1+\cos x)^{2/3}(\sin x)^{1/3}}{\Upsilon^2(1+\cos x)^{4/3}+(2\sin x)^{2/3}}dx,
\ee
where  $"0"$  denotes  an  initial  value  and $x\equiv\frac{\chi}{f}$.\newline
Fortunately, the above integral can be evaluated analytically which we will come back to it in studying the parameter space, next section.
\vskip 0.5 cm

\subsection*{Chromo-natural vs Gauge-flation}
As mentioned before, in the region of large initial axion values $\chi_0/f\simeq\pi$, the chromo-natural model is reduced to the gauge-flation model \cite{Maleknejad:2011jw,Maleknejad:2011sq}
\bea\label{g-f-model}
S&=&\int d^4x\sqrt{-\verb"g"}(-\frac{R}{2}-\frac14F^a_{~\mu\nu}F_a^{~\mu\nu}+\frac{\kappa}{384}(F^a_{~\mu\nu}\tilde F_a^{~\mu\nu})^2),\nonumber
\eea
where $\tilde F_a^{~\mu\nu}\equiv\epsilon^{\mu\nu\lambda\sigma}F^a_{~\lambda\sigma}$. In fact, considering slow-roll inflation and putting $\psi_{min}$ form \eqref{psisl} into \eqref{U-eff}, we find that $\chi$ with an initial field value larger than $f\pi/2$ is on the minimum of its effective potential. Then, it is possible to integrate out the axion field which in the limit of the large initial axion field values $\chi_0\simeq f\pi$ and upon using the relation $\kappa=3\frac{\lambda^2}{\mu^4}$ reduces to gauge-flation model \cite{SheikhJabbari:2012qf}.
In the above we focused on the regime in
which the axion field is massive while the gauge field is light. Another interesting regime is where the opposite situation happens. That is, gauge fields become
massive while the axions are light which happens in the limit $\gamma\gg1$. This analysis has been performed in \cite{Dimastrogiovanni:2012st} and it has been shown that integrating out the gauge fields, chromo-natural model reduces to a theory which only involves axions and has a K-inflation form action with $(\partial\chi)^4$-type additions. Recently, the stability study of the chromo-natural model has been carried out in \cite{Dimastrogiovanni:2012ew}.

\begin{figure}[hb]
\begin{center}
\includegraphics[angle=0, width=70mm, height=60mm]{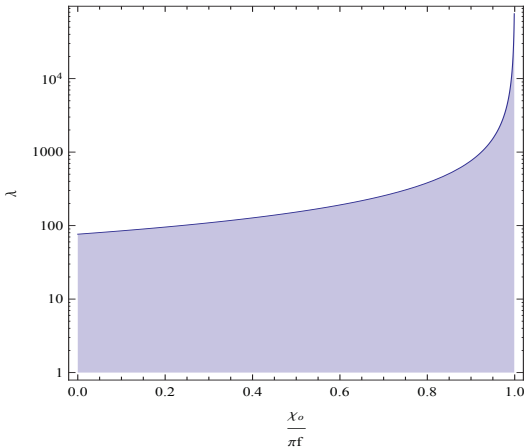}
\caption{In the inequality \eqref{ubn}, we put $N_e=60$ and plotted $\frac{\chi_0}{\pi f}$ vs. $\lambda$. The shaded area, then represents regions in the $\lambda$-$\frac{\chi_0}{\pi f}$ plane that does not lead to a sufficient slow-roll inflation. The white regions is hence suitable for involving inflationary models. }\label{shaded}
\end{center}
\end{figure}

\section{Exploring the parameter space}\label{parameter space}
In this section, upon determining the explicit analytical form of the number of e-folds $N_e$, we find regions in the parameter space which leads to slow-roll inflation with enough number of e-folds.

\begin{figure*}[t] 
   \centerline{\psfig{file=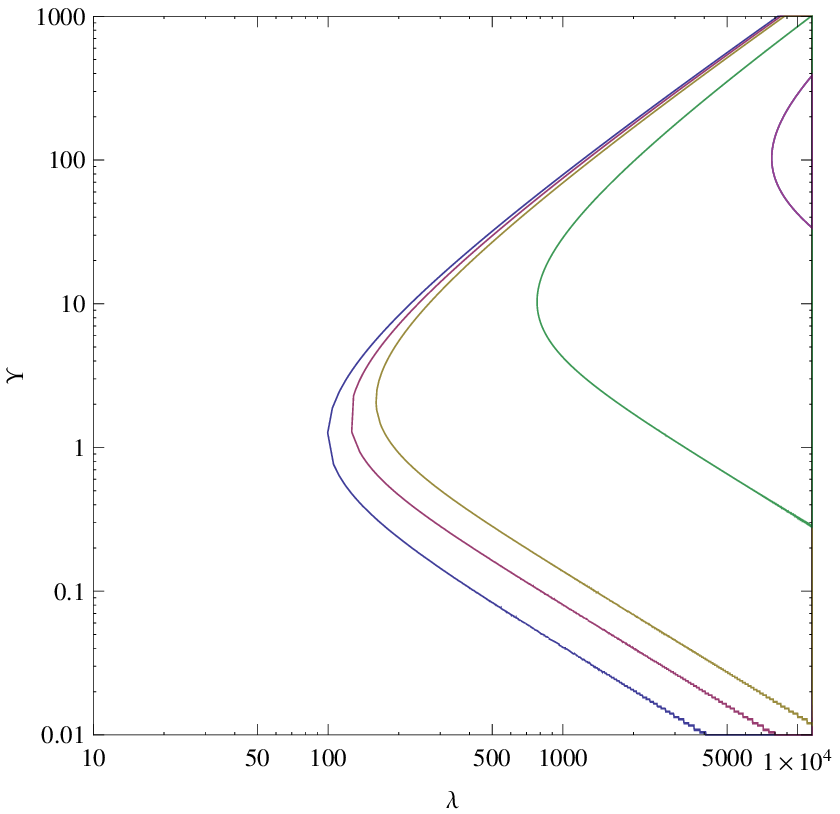, width=3in} \psfig{file=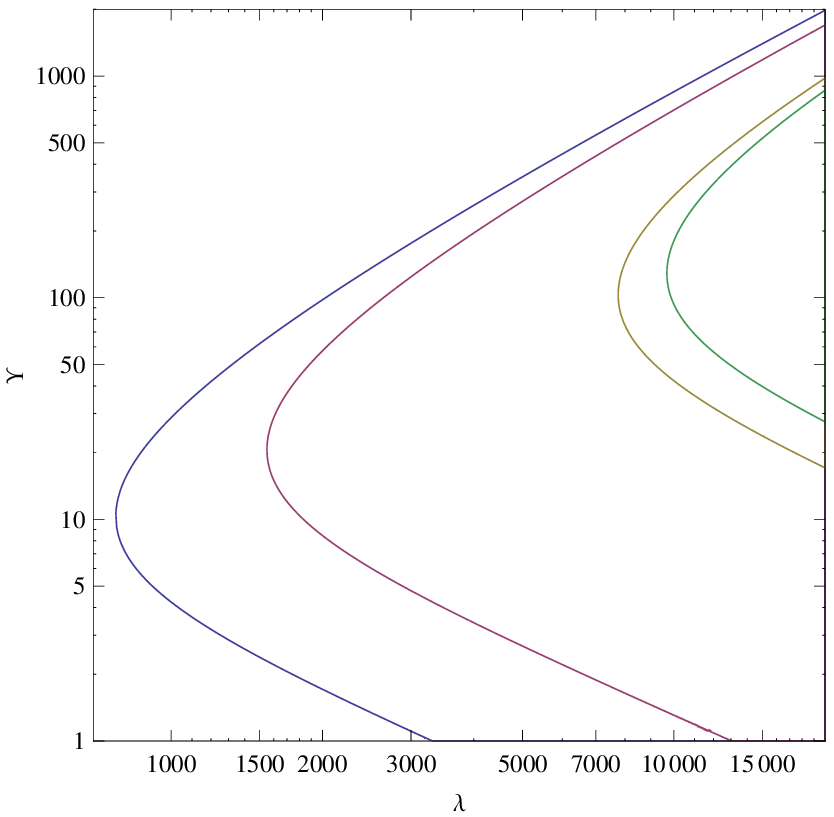, width=3in}}
   \caption{These figures represent the allowed regions of parameter space in the $\lambda-\Upsilon$ plane for $N_e=60$. The curves on the Left figure correspond to various initial values of axion field $\chi_0$; from left to right they correspond to $\frac{\chi_0}{f}=10^{-2}\pi,\ \pi/3, \ \pi/2, \ 0.9\pi$ and $0.99\pi$. The Right plot shows the region on the parameter space which chromo-natural model is effectively equal to the gauge-flation, and the curves from left to right corresponds to $\frac{\chi_0}{f}=0.9\pi, \ 0.95\pi\ , 0.99\pi$ and $0.992\pi$. The ``outer region of the curves correspond to $N_e<60$ and are hence excluded. Note that $\lambda$ is typically (e.g. the minimum value of $\lambda$) of order $10^2$ for the ``small axion values'' (in the left figure), while it is of order $10^4$ for the  large axion values, in the right figure.
The plots imply that $\log(\frac{\lambda}{\lambda_{_{min}}})\geq(\log(\frac{\Upsilon_{_{min}}}{\Upsilon}))^2$, where  $\lambda_{min}=\frac{4\times 60}{\pi-\chi_0/f}$ and $\Upsilon_{_{min}}(\frac{\chi_0}{f})\sim 10^{-2}\lambda_{_{min}}(\frac{\chi_0}{f})$.}
   \label{parameterSpace}
\end{figure*}

Before giving the complete analytic answer of \eqref{N}, it is useful to note that introducing
$$
X\equiv \Upsilon(1+\cos\frac{\chi}{f})^{2/3}\,,\qquad \textmd{and}\quad Y\equiv (2\sin\frac{\chi}{f})^{1/3},
$$
the integrand takes the form $\frac{XY}{X^2+Y^2}$, which is always less than $1/2$ and hence
\be\label{ubn}
N_e\leq\frac{\lambda}{4}(\pi-\frac{\chi_0}{f}).
\ee
The above relation depends only on $\lambda$ and not $\Upsilon$. Hence, demanding $N_e \sim60$ in the inequality \eqref{ubn}, for the hilltop-type small $\chi_0$ case \cite{Adshead:2012kp} we should have $\lambda\sim100$. On the other hand, for the gauge-flation case in which $\pi-\frac{\chi_0}{f}=\mathcal{O}(\epsilon)$, we need larger $\lambda$ values, $\lambda\sim10^4$ \cite{SheikhJabbari:2012qf}. The simple relation \eqref{ubn}, is very useful in restricting the parameter space of the model. Regardless of the values of other parameters, the shaded area in Fig.\ref{shaded} does not lead to a successful slow-roll inflation with enough number of e-folds.


\begin{figure*}[t] 
   \centerline{\psfig{file=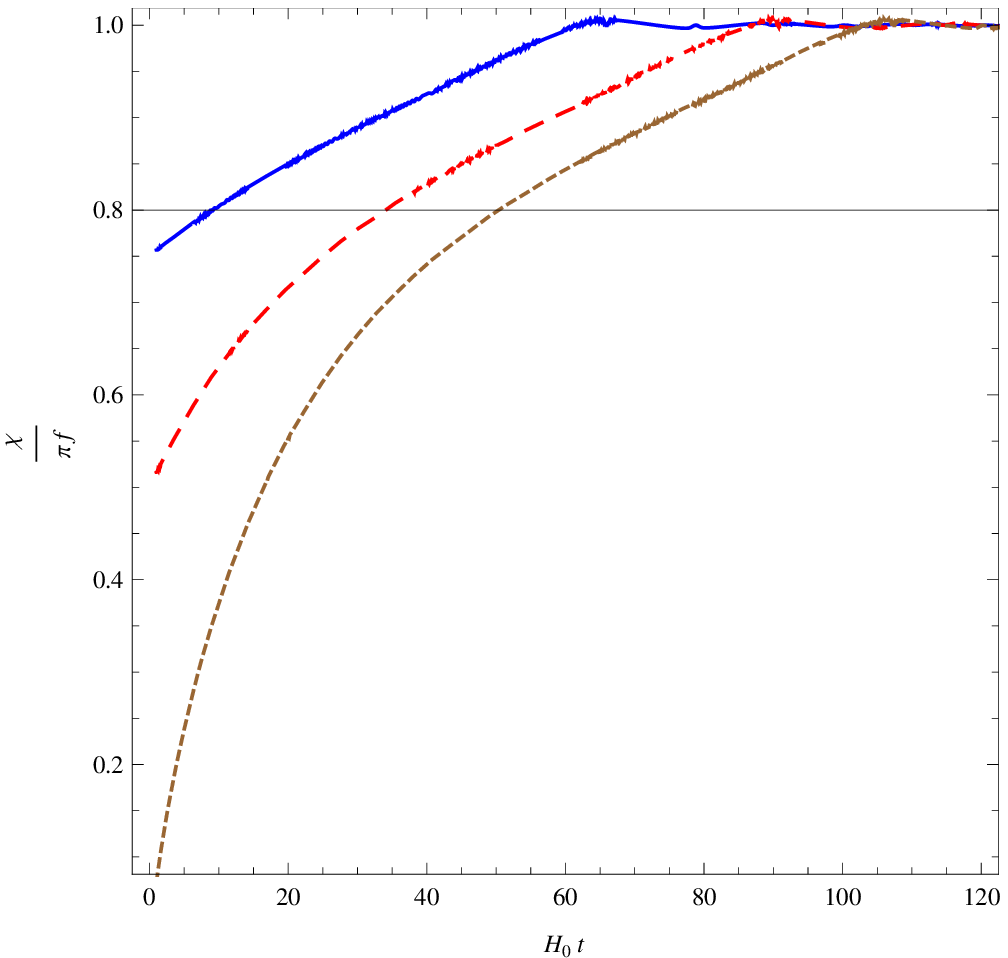, width=2.8in} \psfig{file=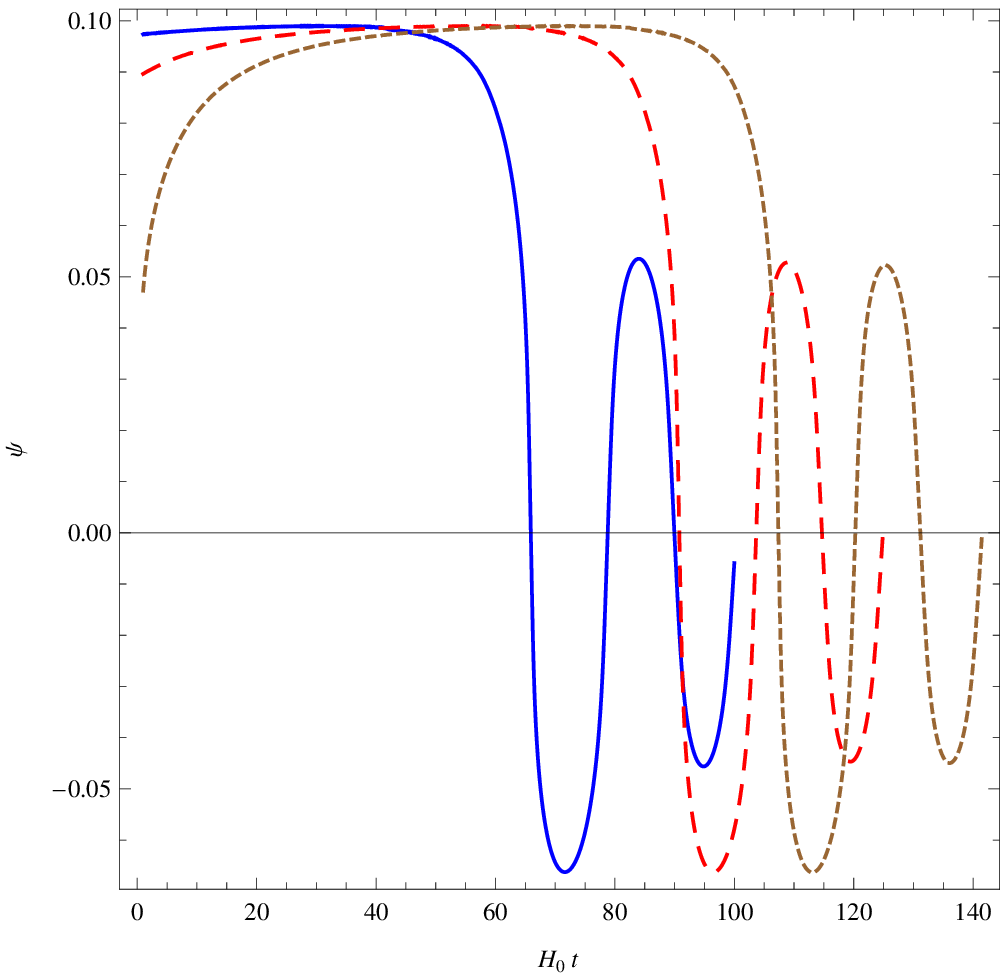, width=2.8in}}
\caption{In this figure we presented the classical trajectories with g=$10^{-6}$, $\lambda$=400, $\mu$=7$\times10^{-4}$, $f=0.009$, started from different axion initial values, $\frac{\chi_0}{f}$. In both panels, the solid (blue) lines, the dashed (red) lines and the dotted (brown) lines correspond to $\frac{\chi_0}{f}$ values equal to $3\pi/4$, $\pi/2$ and $10^{-2} $, respectively. In the left panel, we see that axion field slowly increases during the inflation, it then becomes equal to $f\pi$ and inflation ends. As we see in the right panel, $\psi$  gradually increases during the inflation and after the inflation ends, it starts oscillating just like the gauge-flation model. Note that as the initial value of the axion field increases, $\delta(\chi_0/f)$ becomes smaller, in agreement with our slow-roll analysis. We also note that,  our analytical calculations show, as we decrease initial value of $\chi$ for the given $(g,\lambda, \mu, f)$ values we get a larger number of e-folds. In particular, as we see, the set we have chosen gives 60 number of e-folds for $\chi_0/f=3\pi/4$, while it gives about $N_e=100$ for $\chi_0/f=10^{-2}$. }
   \label{delta-chi}
\end{figure*}

Fortunately, the integral of number of e-folds \eqref{N} can be solved analytically. Introducing
$y\equiv\sin^{2/3}(\frac{\chi}{2f})$, we can write it in the following simple form
\be\label{N-e-exact-chromo-natural}
N_e\simeq \frac32\Upsilon\lambda\int^{1} _{y_0}\frac{y}{\Upsilon^2(1-y^3)+y}dy\,,
\ee
where $y_0={\sin(\frac{\chi_0}{2f})}$. As a third order algebraic equation, the denominator has three roots, one of which is always real and positive $\bar y$, and satisfies
\be\label{ybar}
\Upsilon^{-2}=\bar{y}^2-\frac{1}{\bar{y}}.
\ee
Since $\Upsilon^2>0$, then $\bar{y}>1$.
The other two roots $y_{\pm}$ are given as
\be
y_{\pm}=-\frac{\bar{y}}{2}(1\mp\sqrt{1-\frac{4}{\bar{y}^3}}),
\ee
and depending on the value of $\Upsilon$,  can be either both negative or both complex numbers:\\
$I)~~~$ In case that $\Upsilon\leq \sqrt[6]{\frac{4}{27}}$ which corresponds to $(\bar{y}\geq4^{\frac13})$, $y_{\pm}$ are both real valued and negative. Hence, $N_e$ is given as
\bea\label{N1}
N_e&\simeq &\frac{3\lambda}{2\Upsilon}\bigg[\left(\frac{y_+\ln(y-y_+)}{(\bar{y}-y_+)}-\frac{y_-\ln(y-y_-)}{(\bar{y}-y_-)}\right)\frac{1}{(y_+-y_-)}\nonumber \\
&+&\frac{\bar{y}\ln(\bar{y}-y)}{(\bar{y}-y_+)(\bar{y}-y_-)}\bigg]^{\sin(\frac{\chi_0}{2f})}_{1}\,.
\eea
$II)~~~$  In the case that $\Upsilon> \sqrt[6]{\frac{4}{27}}$ which corresponds to $(\bar{y}\leq4^{\frac13})$, $y_{\pm}$ are  complex valued, $y_{\pm}=-y_R\pm iy_I$, with $y_R=\frac{\bar y}{2}$ and $y_I=\frac{\bar y}{2}\sqrt{\frac{4}{\bar y^3}-1}$. We then obtain the following form for $N_e$
\bea\label{N2}
N_e&\simeq &\frac{3\lambda}{2\Upsilon}\frac{4\bar y^2}{5\bar y^3+16}\bigg[\frac{1}{2}\ln((y+y_R)^2+y_I^2)+\ln(\bar{y}-y) \nonumber \\
&-&(\frac{3\bar y}{4y_I}+\frac{y_I}{\bar{y}})\arctan(\frac{y+y_R}{y_I})\bigg]^{\sin(\frac{\chi_0}{2f})}_{1}\,.
\eea
Having the above analytical expressions for $N_e$ as a function of $\Upsilon$, $\lambda$ and $\chi_0/f$, we are now ready to probe the inflationary parameter space in the entire range of $\chi_0/f\in(0,\pi)$.

For solving the flatness and horizon problems, inflation should have lasted for a minimum number of e-folds $N_e$ which depends on the scale of inflation and somewhat to the details of physics after the end of inflation \cite{Inflation-Books}, it is usual to consider $N_e\gtrsim 60$. Setting $N_e>60$ in \eqref{N1} and \eqref{N2}, we can determine the allowed regions of parameter space corresponding to each initial value of axion field, $\chi_0/f\in(0,\pi)$, which leads to successful inflation. Doing so, in Fig.\ref{parameterSpace} we present the parameter space, $\lambda$ versus $\Upsilon$ for several $\chi_0/f$ values between $10^{-2}$ to $0.99\pi$.
Since the outer area of the curves correspond to $N_e<60$, these areas are automatically excluded. As has been determined by our preliminary relation \eqref{ubn} and also confirmed here, the minimum possible value of $\lambda$ is of the order $10^2$ which corresponds to ``hilltop-type model" \cite{Adshead:2012kp}, while it is of order $10^4$ for the large axion values that effectively lead to ``gauge-flation" \cite{SheikhJabbari:2012qf}. In fact, the allowed regions of parameter space is tightly related to the initial value of the axion field $\chi_0$ and  as we move from $\chi_0/f\ll1$ to $\chi_0/f\simeq\pi$, the necessary values of $\lambda(\chi_0/f)$ and $\Upsilon(\chi_0/f)$ are increasing. As indicated in Fig. \ref{parameterSpace}, the accessible regions which lead to successful slow-roll inflation can be described by the following relation
\bea
\log(\frac{\lambda(\frac{\chi_0}{f})}{\lambda_{_{min}}(\frac{\chi_0}{f})})\geq(\log(\frac{\Upsilon_{_{min}}(\frac{\chi_0}{f})}{\Upsilon(\frac{\chi_0}{f})}))^2,
\eea
where $\lambda_{min}=\frac{4\times 60}{\pi-\chi_0/f}$ and $\Upsilon_{_{min}}(\frac{\chi_0}{f})\sim 10^{-2}\lambda_{_{min}}(\frac{\chi_0}{f})$.
We note that $\Upsilon_{_{min}}(\frac{\chi_0}{f})$ is \emph{not} the minimum value of $\Upsilon$, but the value of $\Upsilon$ corresponds to $\lambda_{min}$.
Moreover, using the above relations and \eqref{H2-sl}, we estimate the scale of inflation in terms of $g$ and $\chi_0/f$
\be
H\sim g\times\frac{\sin^2(\frac{\pi-\chi_0/f}{2})}{(\pi-\chi_0/f)},
\ee
which implies that for a given $H$, the necessary $g$ value is increasing as we move to the large initial axion field values. For instance, setting $H\simeq10^{-5}$, gives $g\sim 10^{-5}$ for the hilltop region, while it gets a more natural value, $g\sim 10^{-3}$, for the region in which the model is effectively gauge-flation. In any case $g$ gives an upper bound for $H$ $$\frac{H}{M_{pl}}\lesssim\frac{g}{4\pi}$$
hence the smallness of $g$ (having a perturbative gauge theory) in the chromo-natural or gauge-flation models is directly associated with the smallness of $H$.

\begin{figure*}[t]
\center{\psfig{file=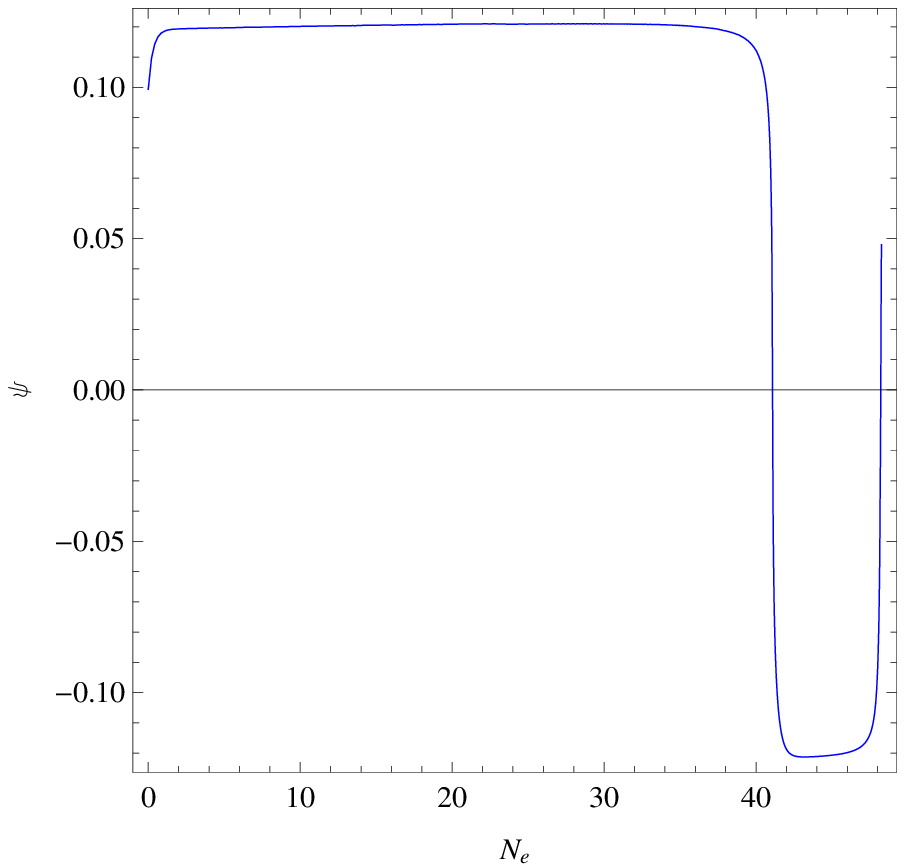, width=2.8in} \psfig{file=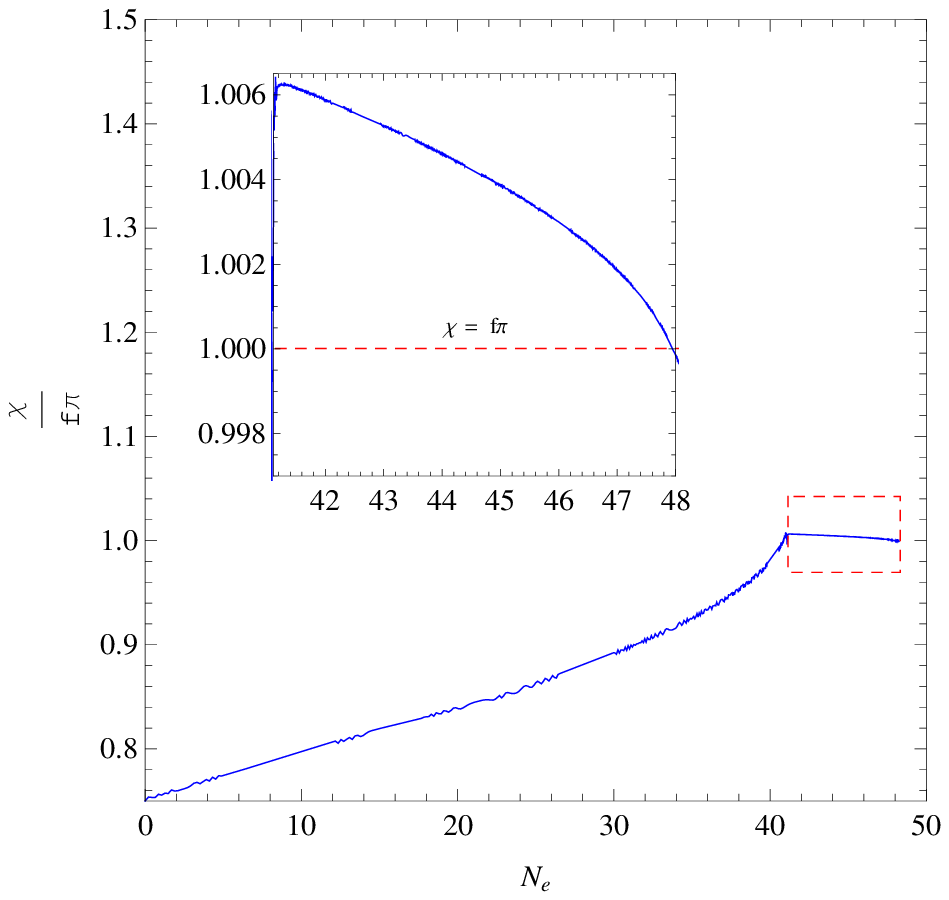, width=3.7in}}
\caption{In this figure we presented a system with g=$10^{-6}$, $\lambda$=3000, $\mu$=2.6$\times10^{-3}$, $f=0.009$ and $\frac{\chi_0}{f}=\frac{3\pi}{4}$ which leads to two phases of slow-roll inflation. In the left panel we have $\psi$ vs. $N_e$, number of e-folding and on the right panel the axion field $\frac{\chi}{f\pi}$ is plotted in terms of $N_e$. This system has two phases, the first one is the conventional slow-roll inflation in which $\psi\simeq\psi_{min}$ and $\chi/f\in(3\pi/4,\pi)$. While the second one is another slow-roll phase which satarted just after the turning point ($\psi=0$ and $\chi=f\pi$), it follows a trajectory with $\psi\simeq-\psi_{min}$ and $\chi>f\pi$. In the right panel, the (red) dashed rectangular shows the behavior of $\frac{\chi}{f\pi}$ during the second slow-roll period which is zoomed in the (red) small box.
Finally, when $\chi=f\pi$ the second stage ends and inflation terminates.}\label{2-slow-roll}
\end{figure*}

\section{Numerical analysis}\label{numerical}
After working out the parameter space which leads to sufficient slow-roll inflation, we are now ready to study the system numerically.  Here, we mainly focus on the behavior of the system with respect to the initial value of the axion field $\chi_0$. For this purpose, after solving the equations \eqref{H2}-\eqref{chieq} and \eqref{psieq} numerically, we present the result for a set of parameters with different $\chi_0/f$ values: $10^{-2}, \pi/2, 3\pi/4$ (Fig.\ref{delta-chi}).

In the left panel of Fig. \ref{delta-chi} we have the time evolution of the axion field which starting from $\chi_0$, it gradually increases during the slow-roll inflation. Then, at the end of inflation $\chi/f$ gets equal to $\pi$ and starts damped oscillations around this value.
The right panel of Fig. \ref{delta-chi} shows $\psi$ field and confirms that during the slow-roll $\psi$ is given by \eqref{psi-sl} and $\delta(\chi_0/f)$ decreases as $\chi_0/f$ gets closer to $\pi$.
In other words, classical trajectories with larger $\chi_0/f$ value, experience smaller $\psi$ field evolution during the slow-roll inflation. Interestingly, regardless of the value of parameters and the initial value of fields, all of these trajectories lead to an oscillatory $\psi$ after the end of slow-roll inflation, as in gauge-flation model.

As a two-field inflationary model, chromo-natural has another interesting feature which is the existence of classical trajectories  with\textit{ two phases} of slow-roll inflation. In cases with large values of $\lambda$ ($\lambda\gg\lambda_{min}(\chi_0/f)$), it is possible to have a second phase of slow-roll inflation in addition to
its conventional slow-roll trajectory. As we present in Fig.\ref{2-slow-roll}, at the transition point:  $\chi=f\pi$ and $\psi=0$.
At this point we have $\dot\chi\gg1$, however from \eqref{chieq} we learn that when $\chi$ is in vicinity of $f\pi$, it has a damped oscillation\footnote{One can write \eqref{chieq} at the transition point as $\ddot{\tilde\chi}+3H\dot{\tilde\chi}+\frac{\mu^4}{f^2}\tilde\chi\simeq0$, where $\tilde\chi\equiv\chi-f\pi\ll1$. At this point, $\dot{\chi}$ has a very large value, however the dynamics of the system leads to a damped osculation of $\tilde\chi$ which quickly decreases $\dot\chi$ just after passing the transition point.} around $f\pi$ which quickly decreases $\dot\chi$. A large $\lambda$ value, then makes it possible that $\dot\chi$ gets very small and leads to the second phase of slow-roll. In the right panel of Fig.\ref{2-slow-roll} we show $\frac{\chi}{f\pi}$ vs. number of e-foldings. The region in the (red) dashed rectangular which corresponds to the second stage of slow-roll is zoomed in a box. As we see, $\dot\chi$ is very large at $f\pi$, but after a short time it gets very small and $\chi$ starts another slow-roll trajectory with $\chi>f\pi$ value. Then, after some e-folds, $\chi$ equals to $f\pi$ and the second phase of slow-roll ends.
At the left panel of Fig\ref{2-slow-roll}, we plotted $\psi$ vs. number of e-folding. The field $\psi$ which was equal to $\psi_{min}$ during the first stage, at the transition point suddenly becomes zero and during the second slow-roll, it gets equal to $-\psi_{min}$, where the minus sign is because $\chi/f>\pi$ in the second stage. We note that the Hubble parameter has different values in the first and the second slow-roll phases.
Such non-trivial background solutions with a transition point may has significant effects on the cosmic perturbation results and would be interesting to investigate.

\section{concluding remarks}\label{conclusion}
Chromo-natural model which was introduced in \cite{Adshead:2012kp} is an interesting inflationary model consisting an axion field and a non-Abelian gauge field. The axion and the gauge field are interacting through a Chern-Simons term which makes it is possible to have successful slow-roll inflation even with the sub-planckian (natural) values of the axion decay constant $(f\ll M_{pl})$. Considering very small initial value for the axion field ($\chi_0/f\ll1$), one has a hilltop-type model associated with the maximum of the axion potential which is the model discussed in \cite{Adshead:2012kp,Adshead:2012qe}. On the other hand, one may choose to start with a large initial axion field value $\chi_0/f\simeq\pi$, then the system is effectively the gauge-flation model which has been discussed in \cite{SheikhJabbari:2012qf}. However, the chormo-natural model has a much broader parameter space to explore. In fact, starting from any arbitrary initial axion field value $\chi_0/f\in(0,\pi)$, it can lead to a successful slow-roll inflation. In this work, we mainly focused on this aspect of the model and working out the analytical solution of number of e-folding, we determined the allowed space of parameters in the $\lambda-\Upsilon$ plane ($\Upsilon\equiv(\frac{2\lambda\mu^4}{3g^2})^{\frac13}$) corresponds to any arbitrary $\chi_0$ value. The parameter space is tightly related to the initial value of the axion field and as we move from $\chi_0/f\ll1$ to $\chi_0/f\simeq\pi$, the allowed values of $\lambda(\chi_0/f)$ and $\Upsilon(\chi_0/f)$ are increasing.  For instance, $\lambda$ is of the order $10^2$ for the small values of $\chi_0$, while it is of the order $10^4$ when $\chi_0$ is in the vicinity of $f\pi$. Moreover, since $\Upsilon$ is proportional to $\mu^2$, as we move from $\chi_0/f\ll1$ to $\chi_0/f\simeq\pi$, $\mu$ is increasing for a constant $g$ value. On the other hand, the parameter $g$ puts an upper bound value on the Hubble parameter during inflation $H\lesssim\frac{g}{4\pi}$, which is a direct relation between the smallness of $g$ (having a perturbative gauge theory) and the scale of inflation.

Next, we investigated the system numerically and studied behavior of the solutions with respect to the initial value of the axion field. Initiating from any arbitrary $\chi_0$ value, axion field has a slowly varying dynamics and when it approaches $f\pi$ inflation ends. Moreover, the slow-roll inflation enforces $\psi$ to be equal to $$\psi_{min}\simeq(\frac{\Upsilon}{\lambda})^{1/2}(\sin(\frac{\chi}{2f}))^{1/3}$$
during the slow-roll, which indicates that classical trajectories with larger $\chi_0/f$ values, experience smaller $\psi$ field evolution during the slow-roll inflation. We note that decreasing $\chi_0/f$ for given $(g,\lambda,\mu)$ values, leads to larger number of e-foldings. Then, in the vicinity of $\chi=f\pi$, slow-roll inflation ends and the field $\psi$ starts oscillating similar to gauge-flation model.

As an interesting feature of chromo-natural which is a two-field inflationary model, we found trajectories with \textit{two phases} of slow-roll inflation. This may have its own significant effects on the cosmic perturbation results which would be interesting to investigate in future works.

\section*{\small Acknowledgment}

It is a pleasure to thank M.~M. Sheikh-Jabbari for his collaboration in the early stages of this work
and for insightful discussions. Work of AM is supported in part by the grands from \emph{Boniad Melli Nokhbegan of Iran}.


\end{document}